\documentclass[journal]{IEEEtran}

\def\figname{Fig.}

\usepackage{cite}
\usepackage{amsmath,amssymb,amsfonts}
\usepackage{algorithmic}
\usepackage{graphicx,color}
\usepackage{textcomp}
\usepackage{url}
\usepackage{float}
\usepackage{bm}
\usepackage{array}
\usepackage{makecell}
\usepackage{flushend}

\def\BibTeX{{\rm B\kern-.05em{\sc i\kern-.025em b}\kern-.08em
    T\kern-.1667em\lower.7ex\hbox{E}\kern-.125emX}}

\begin{document}

\title{Recurrent Transformer-Based Near- and Far-Field THz Wideband Channel Estimation for UM-MIMO}

\author{%
\IEEEauthorblockN{Dmitry Artemasov$^{1}$,
Alexander Shmatok$^{1}$,
Kirill Andreev$^{1}$,
Alexey Frolov$^{1}$,\\
Manjesh K. Hanawal$^{2}$,
and Nikola Zlatanov$^{3}$}\\[2pt]
\IEEEauthorblockA{$^{1}$Center for Next Generation Wireless and IoT, Skolkovo Institute of Science and Technology, Moscow, Russia\\
$^{2}$Department of IEOR, Indian Institute of Technology Bombay, India\\
$^{3}$Faculty of Computer and Engineering Sciences, Innopolis University, Innopolis, Russia\\
Corresponding author: Dmitry Artemasov (d.artemasov@skoltech.ru)}%
\thanks{This work was presented in part at the IEEE SIBIRCON 2024 conference~\cite{artemasov2024sibircon} [DOI: 10.1109/SIBIRCON63777.2024.10758487]. This work was supported by Russian Science Foundation (\protect\url{https://rscf.ru/en/project/24-41-02023/}) under Grant 24-41-02023. The work of Manjesh~Kumar~Hanawal was supported by the Indo-Russian Collaborative Research Funds from Department of Science and Technology (DST).}}

\maketitle

\begin{abstract}
The integration of terahertz communications and ultra-massive multiple-input multiple-output (UM-MIMO) systems in 6G networks is motivated by their ability to enable unprecedented data rates, mitigate spectrum congestion, and enhance overall network performance. However, the enlarged antenna apertures and higher carrier frequencies in these systems increase the Rayleigh distance, causing users to span both the near-field and conventional far-field regions. Accurate spatial precoding thus requires exact channel estimation at the base station -- a task made more challenging by the hybrid coexistence of near- and far-field effects and the limited number of digital chains available in hybrid beamforming architectures. In this paper, we propose a block recurrent transformer model to address this challenge. We demonstrate that a single transformer block equipped with state memory can be trained once and then iteratively applied for hybrid-field channel estimation. Furthermore, we train the model such that it generalizes to wireless channels with varying scatterer distances, different numbers of propagation paths, and wideband operation. Simulation results show that the proposed method achieves performance gains of approximately 5 dB and 7.5 dB in normalized mean squared error (NMSE) over state-of-the-art solutions in narrowband and wideband scenarios, respectively.
\end{abstract}


\section{Introduction}
\label{sec:intro}

\IEEEPARstart{W}{ireless} communication networks face increasing challenges due to the rising demand for higher data rates, greater network density, and improved reliability. To meet these demands, the transition to sixth-generation (6G) networks is expected by 2030~\cite{2024ITU-IMT2030}. Key enabling technologies for 6G include machine learning (ML)-based signal processing, which supports ML approaches to tasks such as channel estimation, signal detection, interference mitigation at the physical layer, and the use of the terahertz (THz) frequency band to support ultra-high-speed links. The adoption of THz frequencies is especially important as lower bands are increasingly congested by existing wireless standards and may lack the bandwidth required to support advanced 6G applications, such as immersive communication~\cite{2024Wei}.

At the same time, the utilization of THz bands presents several challenges. In this frequency range, transmitted signals suffer from severe attenuation due to propagation and atmospheric absorption losses~\cite{2011Akyildiz}. To address this issue, ultra-massive multiple-input multiple-output (UM-MIMO) systems can be employed to concentrate energy in the desired direction using highly directional beams. However, as the number of antenna elements increases, placing radio frequency (RF)-chains at each antenna becomes increasingly difficult due to the limitations of existing hardware.

Hybrid analog-digital beamforming (HBF) has emerged as an effective solution for implementing UM-MIMO systems, as it reduces the number of expensive and power-intensive digital RF-chains. Hybrid beamforming, particularly with partially connected RF-chains architectures, offers a more scalable and power-efficient alternative compared to fully connected architectures, enhancing feasibility for real-world applications~\cite{2024HybridBeamforming}. Nonetheless, HBF architectures introduce new challenges, such as the need for specialized channel estimation algorithms tailored to their unique structure.

Traditional massive MIMO systems typically operate in the sub-6 GHz frequency bands, where signal propagation primarily occurs in the far-field region. However, with the introduction of millimeter-wave (mmWave) and especially THz UM-MIMO technologies, near-field propagation effects become increasingly relevant~\cite{2024Renzo}. The boundary between the near-field and far-field regions is defined by the Rayleigh distance, which depends on both the aperture of the antenna array and the carrier frequency~\cite{2017Krishnasamy}. For instance, a system with an antenna array aperture of 0.5 meters operating at a carrier frequency of 0.1 THz exhibits a near-field region extending approximately up to 200 meters.

Various methods for near-field channel estimation have been well studied in the literature. For example, \cite{2018Tsai, 2020Han, 2021Dovelos, 2011Cai} propose low-complexity algorithms based on compressed sensing (CS) techniques, such as orthogonal matching pursuit (OMP), to address the problem of sparse signal recovery. Usually, in the OMP approach, the search for array response vectors is performed over the predefined discrete Fourier transform (DFT) codebook, which is suitable for the far-field scenarios but causes leakage when applied to near-field channels. As an advancement of the codebook design for the OMP algorithm, the work~\cite{2022Dai} proposes a polar codebook, which samples the space in both angular and distance domains, maintaining sparsity of the channel representation and improving the channel estimation performance. Later, authors of~\cite{2025Liu} further advanced this idea by applying deep learning (DL) techniques to design unique angular-distance codebooks to fit the specific base stations (BSs) deployment scenarios. 
Complementary to codebook-based CS,~\cite{2025LiMadhukumar} proposes learning an adaptive sparsifying dictionary via a batch-delayed online dictionary-learning (BD-ODL) algorithm matched to a practical hybrid-field channel model (including molecular absorption and reflection attenuation).
Non-CS-based methods of near-field channel estimation are also considered in several works. In~\cite{2023Bjornson}, authors propose to perform the near-field channel estimation by first applying the multiple signal classification (MUSIC) algorithm to estimate both the distance and angle of the signal source, and later reconstruct the channel from the estimated parameters. The algorithms in the described papers achieve good accuracy of channel estimation in some near-field scenarios, but their performance may be decreased in other scenarios, especially when the channel is a mixture of near- and far-field components. In parallel, contemporaneous directions include tensor estimators for dynamic metasurface antennas (DMA)-based extremely large (XL)-MIMO wideband systems~\cite{2025ZhangTensor} and THz unmanned aerial vehicle (UAV) integrated sensing and communications (ISAC) transceiver perspectives~\cite{2025ZhangUAV}.

It is important to note that in real-world scenarios, the wireless channel typically includes multiple propagation paths, which may result in the superposition of near-field and far-field components at the receiver, depending on the distances to the scatterers~\cite{2025Wang}. In the following discussion, we refer to the channel containing both near- and far-field components as a hybrid-field channel. Several works describe the application of deep learning techniques for such hybrid-field channels. In \cite{2025Nie}, the authors propose a deep learning-based beam training method that does not rely on predefined beam codebooks. Thereby, instead of searching for the best beam from a pre-fixed codebook, the system learns to directly output the optimal beamforming vector using a convolutional neural network (CNN) trained on historical channel state information (CSI). In~\cite{2024Lei}, the authors adapted the multiple residual dense network (MRDN) model for hybrid-field channel estimation by incorporating a polar-domain transform from~\cite{2022Dai}. This resulted in the polar-domain multi-scale residual dense network (P-MSRDN), which improved channel estimation accuracy. Authors of \cite{2023Zhang} propose to construct a spatial gridding-based sparsifying dictionary and apply the learning iterative shrinkage and thresholding algorithm (LISTA) to achieve accurate channel estimation by leveraging the sparse nature of the channel and thereby overcome the limitations of conventional CS methods. Next, \cite{2023Zhang} proposes a sparsifying dictionary learning LISTA (SDL-LISTA) algorithm to solve the problem of increasing computational complexity, while achieving high channel estimation accuracy by optimizing the dictionary in conjunction with the channel estimator. The work in \cite{2022Letaief} introduces a fixed point network (FPN) framework for hybrid-field channel estimation, addressing the challenge of jointly capturing far- and near-field propagation effects. The proposed channel estimation framework in \cite{2022Letaief} combines a classic closed-form linear estimator based on orthogonal approximate message passing (OAMP) with a deep learning CNN-based non-linear estimator. 

Building on the growing role of deep learning in hybrid-field channel estimation, there has been increasing interest in leveraging more advanced neural network architectures to improve performance and efficiency. One such architecture is the transformer neural network, first introduced in 2017~\cite{2017Vaswani}. Originally developed for natural language processing (NLP), transformers have since been applied in image classification~\cite{2021Dosovitskiy}, time series forecasting~\cite{2022Zeng}, and signal processing~\cite{2024Artemasov}. Relying on an attention mechanism, transformers effectively capture complex dependencies in input data. In wireless communications, recent works have begun to exploit these advantages for UM/THz MIMO: a transformer-based parametric channel acquisition method (T-PCA) learns the mapping from pilot signals to sparse geometric channel parameters (angles, ranges, and gains), enabling channel reconstruction~\cite{Kim2023}. In another work~\cite{Li2024}, a mixed-attention transformer channel estimation network (MAT-CENet) combines feature-map and spatial attention in a transformer encoder to directly estimate near-, far-, and hybrid-field channels, outperforming LS/LMMSE and compressed-sensing baselines. Compared to classical recurrent networks (RNNs) based on recurrent units, such as gated recurrent units (GRU) and long short-term memory (LSTM), transformers often require fewer computational resources and less training time to achieve similar accuracy~\cite{2019Karita}. To combine the strengths of attention and recurrence, Hutchins et al.~\cite{2022Hutchins} introduced the block-recurrent transformer (BRT), which applies a transformer layer recurrently over blocks of tokens and was originally validated on long-sequence language-modeling benchmarks. In this paper, we adopt the BRT architecture for hybrid-field channel estimation -- marking, to the best of our knowledge, the first application of BRT~\cite{2022Hutchins} in a wireless signal processing context.

By adopting the BRT architecture and introducing new techniques for training the corresponding model, we present a novel iterative deep learning framework for hybrid-field channel estimation that improves estimation accuracy compared to previous works and exhibits good generalization across diverse channel conditions. The proposed method iteratively applies a transformer with hidden state memory, such that with each iteration, the channel estimation accuracy improves. Experimental evaluations demonstrate that the proposed approach outperforms both conventional and state-of-the-art neural network-based methods for hybrid-field channel estimation described in the literature in terms of accuracy and performance.\footnote{This paper builds on our conference version~\cite{artemasov2024sibircon}, where the BRT model was initially applied. Here, we advance that study by enhancing generalization and extending the approach to wideband channels, as will be shown in the numerical results.}

The main contributions of this paper are as follows:\\

\begin{itemize}
    \item \textbf{Introduction of the BRT architecture for channel estimation.}\footnote{We note that the BRT architecture differs from the conventional transformer and has not previously been applied in wireless communications.} A novel transformer-based model is proposed for hybrid-field channel recovery in THz UM-MIMO systems, demonstrating superior performance over state-of-the-art approaches in the literature.
    \item \textbf{Constant-parameter BRT for UM-MIMO channel estimation.} The proposed iterative deep-learning framework uses a single transformer block with state memory, applied recurrently to refine hybrid-field estimates while keeping the total number of trainable parameters fixed regardless of the number of refinement iterations.
    \item \textbf{Robust generalization across channel conditions.} The proposed model demonstrates strong adaptability under varying scatterer distributions, number of multipath components, and bandwidths, ensuring reliable performance across a broad range of deployment scenarios.
    \item \textbf{Effective wideband channel estimation.} By treating orthogonal frequency-division multiplexing (OFDM) subcarriers as sequential tokens, the proposed framework captures spectral dependencies between the subcarriers using multi-head self-attention and recurrence, enabling accurate estimation in wideband hybrid-field THz UM-MIMO systems.
    \item \textbf{State-of-the-art performance.} The proposed method achieves significant improvement over the state-of-the-art hybrid-field deep learning-based solution (FPN-OAMP)~\cite{2022Letaief} across the $0$--$20$~dB signal-to-noise ratio (SNR) range in both narrowband and wideband channel settings. Specifically, around $5$~dB and $7.5$~dB improvement in normalized mean squared error (NMSE) is observed in narrowband and wideband scenarios, respectively.
\end{itemize}

We note that the proposed estimator is channel model-agnostic and can be applied for massive MIMO channel estimation in various scenarios, including purely far-field and sub-THz bands. However, in this paper, we adopt the hybrid-field THz UM-MIMO wideband setting as a stringent benchmark. This regime jointly stresses estimator design due to: (i) mixed near-/far-field propagation that induces range--angle coupling and transitions between spherical and planar wavefronts; (ii) wideband effects (e.g., beam squint and frequency-selective dispersion); and (iii) hybrid-beamforming hardware constraints (few RF chains, constant-modulus/quantized phase control, and limited pilot budgets) that reduce instantaneous observability. Together, these factors make the channel estimation problem more ill-conditioned than purely far-field, narrowband, or fully-digital scenarios, thereby providing a challenging and discriminative testbed. We therefore compare BRT against strong, regime-appropriate baselines in this setting. The architecture itself remains general and, with the appropriate system model and training data, can be applied to other massive MIMO systems.\footnote{We release our simulator, data generation scripts, and BRT implementation at \url{https://github.com/dartemasov/brt-channel-estimation} to enable directly comparable evaluation under identical conditions.}

The rest of the paper is organized as follows. In Section~\ref{sec:system-model}, the system model is introduced. In Section~\ref{sec:problem-statement}, the channel estimation problem is formulated. In Section~\ref{sec:channel-estimation}, the BRT model architecture and its adaptation for the channel estimation task are described. Section~\ref{sec:training-eval-setup} provides the parameters of channel simulation and the proposed block recurrent transformer model training setup. Section~\ref{sec:robustness} addresses the model's generalization ability to the hybrid-field channels with varying parameters. In Section~\ref{sec:wideband}, the operation of the proposed model in the wideband scenario is described. Finally, Section~\ref{sec:conclusions} concludes the paper and discusses the potential research directions.

\section{System Model}
\label{sec:system-model}

To enable straightforward performance comparison with state-of-the-art results, we adopt the UM-MIMO hybrid-field THz system and channel models from~\cite{2022Letaief}.

\begin{figure}
    \centering
    \includegraphics[width=\columnwidth]{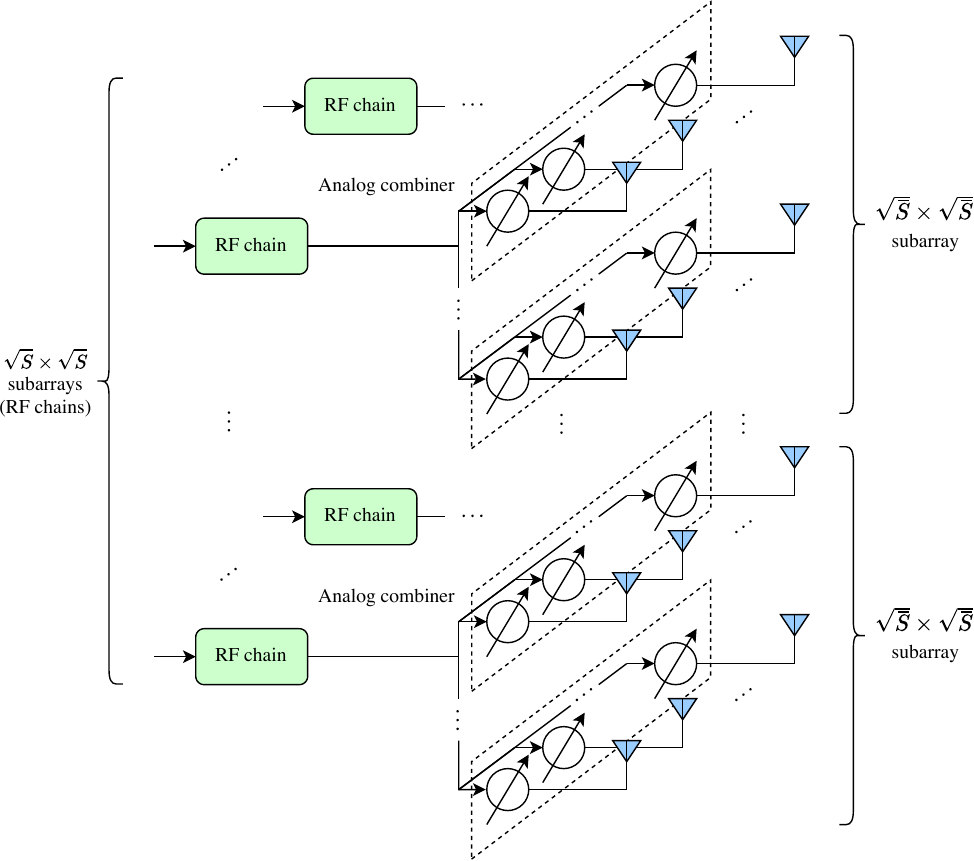}
    \caption{Partially connected hybrid beamforming structure: each AE in the subarray is connected to a single RF chain via a phase shifter.}
    \label{fig:AoSA}
\end{figure}

The multi-antenna BS is assumed to employ the HBF architecture, specifically the array-of-subarrays (AoSA) architecture~\cite{2022Letaief}. Thereby, the system model is comprised of a multi-antenna BS and single-antenna user equipments (UEs). Hence, the BS consists of a $\sqrt{S} \times \sqrt{S}$ grid of antenna subarrays (SAs), where each SA contains $\sqrt{\bar{S}} \times \sqrt{\bar{S}}$ antenna elements (AEs). Next, each SA has a single dedicated RF chain, whereas each AE within the SA is connected to the respective RF chain via an analogue phase shifter. A schematic illustration of such a HBF architecture is provided in \figname~\ref{fig:AoSA}.

Following~\cite{2022Letaief} and~\cite{2020Letaief}, we define the index of the SA located at the $m$-th row and $n$-th column of the AoSA as $s=(m-1)\sqrt{S}+n$, where $1\leq m,n\leq\sqrt{S}$ and $1\leq s\leq S$. Similarly, for a given SA, the index $\bar{s}$ of the AE positioned at the $\bar{m}$-th row and $\bar{n}$-th column within the given SA is determined by $\bar{s} = (\bar{m}-1)\sqrt{\bar{S}}+\bar{n}$, where $1\leq\bar{m}, \bar{n}\leq\sqrt{\bar{S}}$ and $1\leq\bar{s}\leq\bar{S}$.

\begin{figure}
    \centering
    \includegraphics[width=0.75\columnwidth]{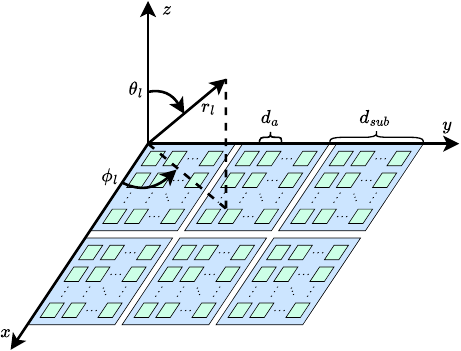}
    \caption{AoSA schematic representation.}
    \label{fig:array}
\end{figure}

The distance between two neighboring SAs is denoted by $d_{\text{sub}}$, see \figname~\ref{fig:array}. Similarly, the distance between two neighboring AEs within a given SA is denoted by $d_{a}$, also see \figname~\ref{fig:array}. As depicted in \figname~\ref{fig:array}, we set up a Cartesian coordinate system with its origin located at the first AE of the first SA. Assuming that the AoSA lies within the $x$-$y$ plane, the position of the $\bar{s}$-th AE in the $s$-th SA, within this coordinate system, can be expressed as
\begin{equation}
\mathbf{p}_{s,\bar{s}}=\left[\begin{array}{c}
(m-1)[(\sqrt{\bar{S}}-1)d_{a}+d_{\text{sub}}]+(\bar{m}-1)d_{a}\\
(n-1)[(\sqrt{\bar{S}}-1)d_{a}+d_{\text{sub}}]+(\bar{n}-1)d_{a}\\
0
\end{array}\right].
\label{eq:ae-position}
\end{equation}

The channel response $\mathbf{\tilde{h}} \in \mathbb{C}^{S\bar{S}}$ between the BS and a single UE is modeled as the superposition of contributions from $L$ propagation paths, and is given by
\begin{equation}
    \mathbf{\tilde{h}}=\gamma\sum_{l=1}^{L}\alpha_{l}\mathbf{a}\left(\phi_{l},\theta_{l},r_{l}\right)e^{-j2\pi f_{c}\tau_{l}},
    \label{eq:channel}
\end{equation}
where $\gamma$ is a normalization factor such that $\lVert \mathbf{\tilde{h}} \rVert_2^2 = S\bar{S}$ holds~\cite{2022Letaief}, $\alpha_l$ denotes the path gain of the $l$-th path, $f_c$ is the carrier frequency, and $\tau_l$ is the propagation delay of the $l$-th path. The vector $\mathbf{a}\left(\phi_{l},\theta_{l},r_{l}\right)$ denotes the array response vector, which will be defined in detail later on. The parameters $\phi_l$, $\theta_l$, and $r_l$ represent the azimuth angle of arrival (AoA), elevation AoA, and distance to the RF signal source of the $l$-th path, respectively. The nature of the RF source depends on the propagation conditions: in LoS scenarios, the UE acts as the RF source, while in NLoS scenarios, the effective RF source is the final scatterer or reflector along the path. Let $Z$ denote the Rayleigh distance, obtained as
\begin{equation}
    Z = \frac{2D^2}{\lambda_c},
\end{equation}
where $D$ is the aperture of the antenna array and $\lambda_c$ is the operating wavelength. Then, if the distance $r_l$ is smaller than the Rayleigh distance $Z$, $\mathbf{a}(\phi_l, \theta_l, r_l)$ corresponds to the near-field response, denoted by $\mathbf{a}^{NF}$; otherwise, it corresponds to the far-field response, denoted by $\mathbf{a}^{FF}$, i.e.,
\begin{equation}
    \mathbf{a}(\phi_l, \theta_l, r_l) = 
    \begin{cases}
        \mathbf{a}^{NF}(\phi_l, \theta_l, r_l) & \text{if } r_l < Z,\\
        \mathbf{a}^{FF}(\phi_l, \theta_l, r_l) & \text{otherwise}.
    \end{cases}
\end{equation}

In the near-field, each element of the response vector $\mathbf{a}^{NF}(\phi_{l}, \theta_{l}, r_l)$ depends on the exact distance between each antenna element and the RF source, due to the spherical wavefront assumption. Thereby, $\mathbf{a}(\phi_{l}, \theta_{l}, r_l) = \text{vec}\big(\mathbf{A}^{NF}(\phi_{l}, \theta_{l}, r_l)\big)$, where~\cite{2022Letaief}
\begin{equation}
\mathbf{A}^{NF}(\phi_{l},\theta_{l},r_{l})_{s,\bar{s}}=e^{-j2\pi\frac{f_{c}}{c}\|\mathbf{p}_{s,\bar{s}}-r_{l}\mathbf{t}_{l}\|_{2}}.
\label{eq:near-field-response}
\end{equation}
In (\ref{eq:near-field-response}), $c$ represents the speed of light, $\mathbf{p}_{s,\bar{s}}$ is defined in equation (\ref{eq:ae-position}) and represents the position of the $\bar{s}$-th AE in the $s$-th SA, and $\mathbf{t}_l$ denotes the unit vector specifying the direction of arrival of the $l$-th propagation path and is given by
\begin{equation}
    \mathbf{t}_{l}=[\sin\theta_{l}\cos\phi_{l},\sin\theta_{l}\sin\phi_{l},\cos\theta_{l}]^{T}.
\end{equation}

The far-field array response vector is similarly defined as $\mathbf{a}^{FF}(\phi_{l},\theta_{l}) = \text{vec}\big(\mathbf{A}^{FF}(\phi_{l},\theta_{l})\big)$, where
\begin{equation}
\mathbf{A}^{FF}(\phi_{l},\theta_{l},r_l)_{s,\bar{s}}=e^{-j2\pi\frac{f_{c}}{c}(\mathbf{p}_{s,\bar{s}}^{T}\mathbf{t}_{l} - r_l)}.
\label{eq:far-field-response}
\end{equation}
Here, $(\mathbf{p}_{s,\bar{s}}^{\mathsf T}\mathbf{t}_l - r_l)$ is the first-order approximation of the propagation distance obtained via a Taylor expansion~\cite{2022Letaief}.

In the THz band, the signal propagating via different paths suffers from molecular absorption loss and reflection loss~\cite{2021Dovelos}. The LoS path suffers only from molecular absorption loss, and thereby, if we fix the LoS path index to $l=1$, then its attenuation $\alpha_1$ can be formulated as~\cite{2019Boulogeorgos}
\begin{equation}
    \alpha_1 = \frac{c}{4\pi f_c r_1}e^{-\frac{1}{2}k_\text{abs}r_1},
\label{eq:los-path-loss}
\end{equation}
where $k_\text{abs}$ is the absorption coefficient.

For the reflection loss of the NLoS paths, we adopt the models from \cite{2007Piesiewicz, 2021Dovelos} and thereby define the reflection loss coefficient of the path $l\in[2,\dots,L]$ as
\begin{equation}
    \Gamma_{l}=\frac{\cos\varphi_{\text{in},l}-n_{t}\cos\varphi_{\text{ref},l}}{\cos\varphi_{\text{in},l}+n_{t}\cos\varphi_{\text{ref},l}}e^{-\left(\frac{8\pi^{2}f_{c}^{2}\sigma_{\text{rough}}^{2}\text{cos}^{2}\varphi_{\text{in},l}}{c^{2}}\right)},
    \label{eq:reflection-loss}
\end{equation}
where $\varphi_{\text{in},l}$ is the angle of incidence for the $l$-th path, and $\varphi_{\text{ref},l} = \arcsin(n_{t}^{-1}\sin\varphi_{\text{in},l})$ denotes the angle of refraction, $n_{t}$ represents the refractive index, while $\sigma_{\text{rough}}$ represents the roughness coefficient of the reflecting material. Then, the attenuation of the $l$-th NLoS propagation path is defined as~\cite{2015Lin}
\begin{equation}
    \alpha_l = |\Gamma_l|\alpha_1.
    \label{eq:nlos-path-loss}
\end{equation}

\section{Problem Statement}
\label{sec:problem-statement}

In this work, we focus on the uplink channel estimation task. By allocating distinct time and/or frequency resources to different UEs, their corresponding channel estimates can be obtained independently. Therefore, without loss of generality, the following discussion focuses on the channel estimation procedure for a single UE, as done in~\cite{2022Letaief}.

We assume that UE transmits $N_p$ pilot signals. The received pilot signal $\mathbf{\tilde{y}}_{p}\in\mathbb{C}^{S}$ at the BS is then given by
\begin{equation}
\mathbf{\tilde{y}}_{p}=\mathbf{\widetilde{W}}_{\text{BB},p}^{H}\mathbf{\widetilde{W}}_{\text{RF},p}^{H}\mathbf{\tilde{h}}s_{p}+\mathbf{\widetilde{W}}_{\text{BB},p}^{H}\mathbf{\widetilde{W}}_{\text{RF},p}^{H}\mathbf{\tilde{n}}_{p}, \forall p\in [1,\dots ,N_p],
\label{eq:complex-model}
\end{equation}
where $\mathbf{\widetilde{W}}_{\text{BB},p} \in \mathbb{C}^{S \times S}$ denotes the digital combining matrix, and $\mathbf{\widetilde{W}}_{\text{RF},p} \in \mathbb{C}^{S\bar{S} \times S}$ is the block-diagonal analog combining matrix, whose elements are independently drawn with equal probability from the set of single-bit quantized values $\frac{1}{\sqrt{\bar{S}}}\{-1, 1\}$ to reduce system complexity and power consumption~\cite{2018He}. The normalization factor $\frac{1}{\sqrt{\bar{S}}}$ ensures that the power constraint is satisfied. It is worth mentioning that both digital $\mathbf{\widetilde{W}}_{\text{BB},p}$ and analog $\mathbf{\widetilde{W}}_{\text{RF},p}$ combiners are known to the base station by design. The channel vector is represented by $\mathbf{\tilde{h}} \in \mathbb{C}^{S\bar{S}}$, $s_p$ is the pilot symbol, and $\mathbf{\tilde{n}}_p \sim \mathcal{CN}(\mathbf{0}, \sigma_n^2 \mathbf{I})$ represents complex additive white Gaussian noise (AWGN).

For simplicity of description, we assume that the pilot symbol $s_p = 1$, and the digital combining matrix is set to identity matrix, i.e., $\mathbf{\widetilde{W}}_{\text{BB},p} = \mathbf{I}$, implying that no receive digital beamforming is applied during pilot signal reception. Under these assumptions, the measurement model in equation~(\ref{eq:complex-model}) simplifies to

\begin{equation}
\mathbf{\tilde{y}} = \mathbf{\widetilde{W}}_{\text{RF}}^H \mathbf{\tilde{h}} + \mathbf{\tilde{n}},
\label{eq:complex-stacked-model}
\end{equation}
\makebox[\linewidth][s]{where $\mathbf{\tilde{y}} = [\mathbf{\tilde{y}}_1,\dots, \mathbf{\tilde{y}}_{N_p}]\in\mathbb{C}^{SN_p}$, $\mathbf{\widetilde{W}}_{\text{RF}}^H = $} \\
\makebox[\linewidth][s]{$[\mathbf{\widetilde{W}}_{\text{RF},1}^H,\dots , \mathbf{\widetilde{W}}_{\text{RF},N_p}^H]
\in\mathbb{C}^{SN_p\times S\bar{S}}$, $\mathbf{\tilde{n}} =$} \\
$ [\mathbf{\widetilde{W}}_{\text{RF},1}^H \mathbf{\tilde{n}}_1,\dots, \mathbf{\widetilde{W}}_{\text{RF},N_p}^H \mathbf{\tilde{n}}_{N_p}]\in\mathbb{C}^{SN_p}$.

In this work, we will apply neural network-based signal processing. For convenience, we express equation~(\ref{eq:complex-stacked-model}) in the real-valued form. To this end, we define $\mathbf{y} = [\text{\ensuremath{\Re}}(\mathbf{\tilde{y}}), \text{\ensuremath{\Im}}(\mathbf{\tilde{y}})]\in\mathbb{R}^{2SN_p}$, $\mathbf{h} = [\text{\ensuremath{\Re}}(\mathbf{\tilde{h}}), \text{\ensuremath{\Im}}(\mathbf{\tilde{h}})]\in\mathbb{R}^{2S\bar{S}}$, $\mathbf{n} = [\text{\ensuremath{\Re}}(\mathbf{\tilde{n}}),\text{\ensuremath{\Im}}(\mathbf{\tilde{n}})]\in\mathbb{R}^{2SN_p}$, and
\begin{equation*}
\mathbf{W}_{\text{RF}}^H=
\begin{bmatrix}
\text{\ensuremath{\Re}}(\mathbf{\widetilde{W}}_{\text{RF}}^H) & -\text{\ensuremath{\Im}}(\mathbf{\widetilde{W}}_{\text{RF}}^H) \\
\text{\ensuremath{\Im}}(\mathbf{\widetilde{W}}_{\text{RF}}^H) & \text{\ensuremath{\Re}}(\mathbf{\widetilde{W}}_{\text{RF}}^H)
\end{bmatrix}\in\mathbb{C}^{2SN_p\times 2S\bar{S}}.
\end{equation*}
Now (\ref{eq:complex-stacked-model}) can be reformulated as
\begin{equation}
    \mathbf{y} = \mathbf{W}_{\text{RF}}^H \mathbf{h} + \mathbf{n}.
    \label{eq:real-model}
\end{equation}

The aim of this work is to estimate $\mathbf{\hat{h}}$ from the observation $\mathbf{y}$, given by (\ref{eq:real-model}). In this paper, we adopt the NMSE both as the objective function for minimization and as the performance metric for evaluation, which is defined as
\begin{equation}
    \text{NMSE} = \mathbb{E} \bigg(\frac{\lVert \mathbf{h} - \mathbf{\hat{h}} \rVert^2_2}{\lVert \mathbf{h} \rVert_2^2} \bigg).
    \label{eq:nmse}
\end{equation}

\section{Block Recurrent Transformer Hybrid-Field UM-MIMO THz Channel Estimation}
\label{sec:channel-estimation}

\subsection{BRT Model Architecture}

Unlike conventional transformer architectures, which process the input sequence using stacked layers of self-attention and feed-forward networks, the BRT introduces recurrence into the architecture. Specifically, a single transformer block with state memory is applied recurrently across time steps. This design keeps the number of trainable parameters constant, while enabling the model to progressively refine its outputs through iterative correction. Another difference is that BRT integrates both vertical (layer-wise) and horizontal (time-step) processing directions, with hidden state updates regulated by gating mechanisms. As a result, BRT combines the contextual modeling strength of transformers with the temporal memory advantages of recurrent networks, making it especially suitable for iterative channel estimation tasks.

\begin{figure*}[t]
    \centering
    \includegraphics[width=\textwidth]{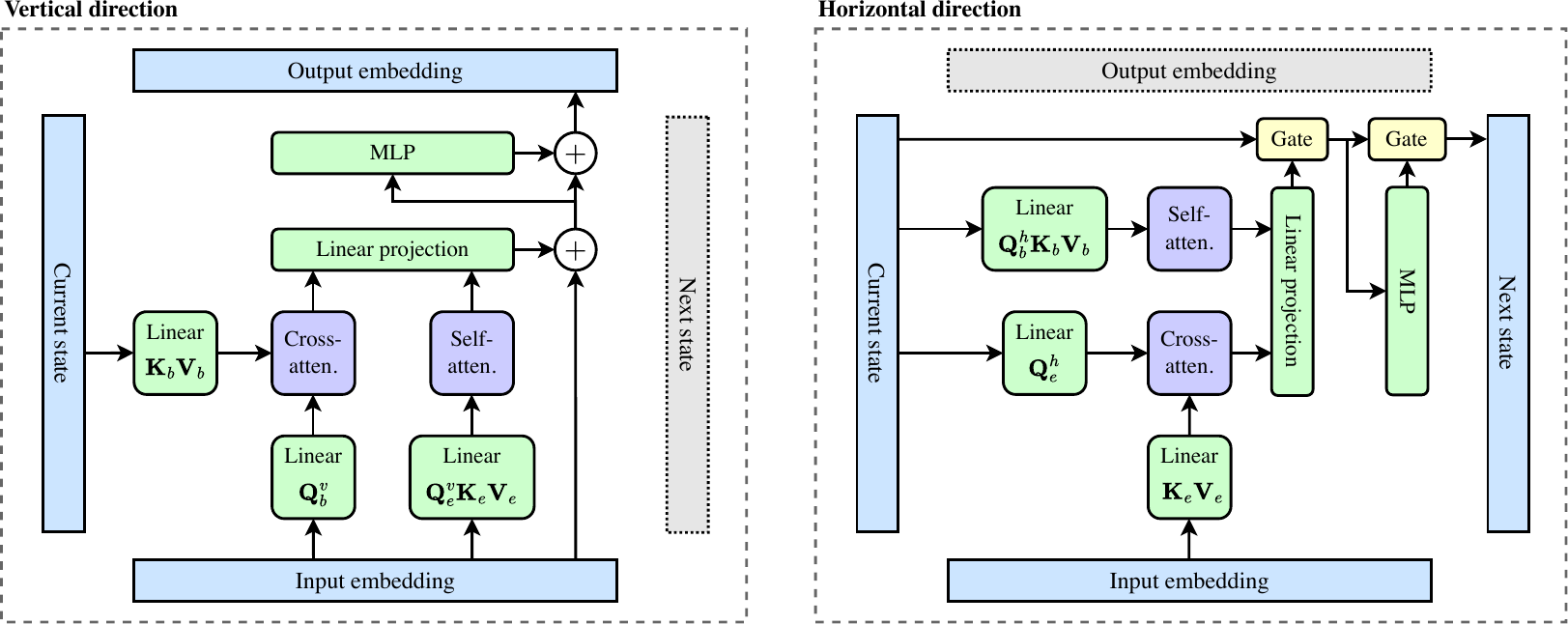}
    \caption{Illustration of the block recurrent transformer cell~\cite{2022Hutchins}. The \textit{vertical} direction is shown on the left: the input embedding and the current hidden state are combined to generate the embedding for the next \textit{vertical} BRT cell. The \textit{horizontal} direction is shown on the right: the next hidden state is computed from the input embedding and the current hidden state.}
    \label{fig:rbt-cell}
\end{figure*}

The BRT model is composed of BRT cells, whose architecture is illustrated in \figname~\ref{fig:rbt-cell}. Each BRT cell takes input token embeddings and the current state as inputs, and produces both output embeddings and next states. BRT cells operate along two dimensions simultaneously: the vertical direction and the horizontal direction, as shown in the left and right panels of \figname~\ref{fig:rbt-cell}, respectively.

In the vertical direction (left panel), the model maps input token embeddings and the current recurrent state to output embeddings. This is achieved using two attention mechanisms:
\begin{itemize}
    \item Self-attention is applied to the input embeddings to capture contextual information.
    \item Cross-attention is applied between the current state and the input embeddings to integrate recurrent information.
\end{itemize}
The results are combined via a linear projection and a multilayer perceptron (MLP), followed by residual connections, to produce the output embeddings.

In the horizontal direction (right panel), the model updates the hidden state for the next time step. This is done by:
\begin{itemize}
    \item Applying self-attention to the current state.
    \item Applying cross-attention between the input embeddings and the current state.
\end{itemize}
The outputs are passed through a linear projection and an MLP, each modulated by gating mechanisms, to produce the next state.

Importantly, the model uses shared keys and values across both directions: $\mathbf{K}_e, \mathbf{V}_e$ from the input embeddings, and $\mathbf{K}_b, \mathbf{V}_b$ from the current state. However, the queries are not shared. Four distinct query sets are used: $\mathbf{Q}_b^v$ and $\mathbf{Q}_e^v$ for the vertical direction, and $\mathbf{Q}_b^h$ and $\mathbf{Q}_e^h$ for the horizontal direction. The described design allows the BRT cell to integrate both temporal (horizontal) and contextual (vertical) information efficiently within each processing step.

An important design feature in the horizontal direction is the presence of a gate mechanism. The gate regulates the cell’s ability to forget or retain information as it propagates across \textit{time steps}. Such controlled information flow has proven crucial for solving algorithmic tasks~\cite{2021Csordas}. In our implementation, we use fixed gates, whose behavior is defined by the following equations
\begin{align}
    \mathbf{z}_t &= \mathbf{U}_z \mathbf{v}_t + \mathbf{s}_z, \\
    \mathbf{g} &= \sigma(\mathbf{s}_g), \\
    \mathbf{c}_{t+1} &= \mathbf{c}_t \odot \mathbf{g} + \mathbf{z}_t (1 - \mathbf{g}),
\end{align}
where $\mathbf{U}_z$ is the trainable weights matrix, $\mathbf{v}_t$ is the gate input (output of linear projection or MLP in our case), $\mathbf{s}_z$ and $\mathbf{s}_g$ are the trainable bias vectors, $\sigma(\cdot)$ is the sigmoid function, $\odot$ denotes the Hadamard product, $\mathbf{c}_t$ and $\mathbf{c}_{t+1}$ are the current and the next cell state vectors, respectively.

\begin{figure}
    \centering
    \includegraphics[width=\columnwidth]{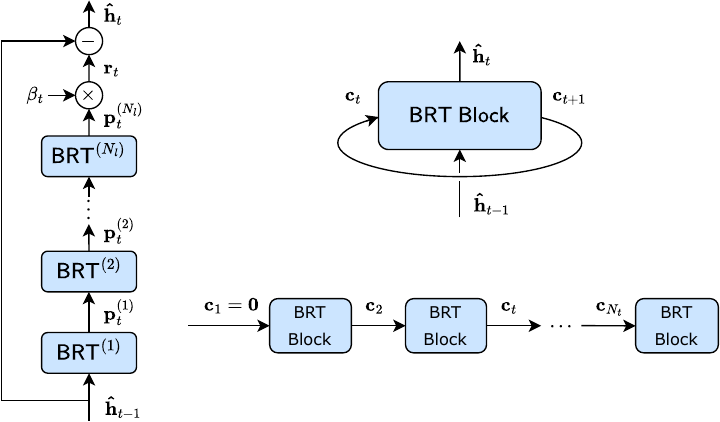}
    \caption{Proposed BRT-based hybrid-field channel estimation model. The vertical processing direction is depicted on the left side, the horizontal processing direction is depicted on the right side.}
    \label{fig:rbt-model}
\end{figure}

\subsection{BRT Model for Channel Estimation}

The BRT model is constructed by stacking BRT cells vertically, forming what we refer to as a \textit{BRT block}. This vertical composition defines the depth $N_l$ of the model, what is schematically depicted on the left side of \figname~\ref{fig:rbt-model}. The cells $[\mathsf{BRT}^{(1)},\mathsf{BRT}^{(2)},\dots,\mathsf{BRT}^{(N_l)}]$ are sequentially arranged, where each BRT layer processes its input, passing the resulting embedding $\mathbf{p}_t^{(l)}$ upward to the next layer.

Once constructed, this BRT block is applied recurrently across $N_t$ \textit{time steps}, as depicted on the right side of \figname~\ref{fig:rbt-model}. At each time step $t$, the BRT block receives as input the current estimation $\mathbf{\hat{h}}_{t-1}$ and the set of memory states $\{\mathbf{c}_t^{(l)}\}_{l=1}^{N_l}$ from the previous step. These are used to compute the output embedding $\mathbf{p}_t^{(N_l)}$, which is scaled by a learnable scalar $\beta_t$ to obtain the residual correction
\begin{equation}
    \mathbf{r}_t = \beta_t\mathbf{p}_t^{(N_l)}.
\end{equation}
This residual is then subtracted from the previous estimate to obtain the refined channel estimate
\begin{equation}
    \mathbf{\hat{h}}_t = \mathbf{\hat{h}}_{t-1} - \mathbf{r}_t.
    \label{eq:res-correction}
\end{equation}

The natural interpretation is that $\mathbf{p}_t^{(N_l)}$ encodes the approximate error in the current estimate $\mathbf{\hat{h}}_{t-1}$. To correct the estimate, the error is removed, i.e., subtracted from the predicted residual, as defined in (\ref{eq:res-correction}). This choice follows the standard residual/iterative-correction convention and makes the role of the transformer output interpretable.\footnote{We note, that some approaches (e.g.,~\cite{2023Bjornson, Kim2023}) first estimate channel parameters -- AoAs, path distances, and gains -- and then reconstruct the channel. In contrast, we estimate the high-dimensional CSI vector directly. Because many baseband algorithms (e.g., beamforming and scheduling) operate on CSI, end-to-end estimation avoids intermediate reconstruction. Empirically, it is also more stable: small errors in multipath-parameter estimates can substantially degrade the final channel estimation quality.}

This recursive refinement is repeated for $t=[1,\dots,N_t]$, where the same BRT block is reused at each time step. As a result, the number of trainable parameters remains constant, independent of $N_t$.

\begin{figure}
    \centering
    \includegraphics[width=\columnwidth]{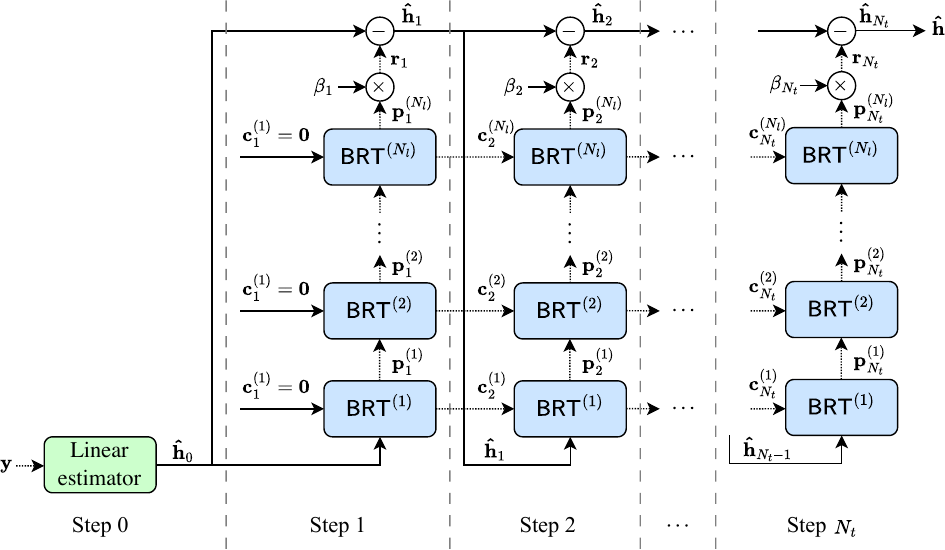}
    \caption{Proposed BRT-based hybrid-field channel estimation model. Unrolled representation.}
    \label{fig:rbt-model-unrolled}
\end{figure}

The overall channel estimation pipeline is presented in \figname~\ref{fig:rbt-model-unrolled}. For clarity of the notations, the recurrent structure of the BRT model is unrolled in this illustration. As shown in the figure, the pipeline begins with a linear estimation step, defined as
\begin{equation}
    \mathbf{\hat{h}}_0 = \mathbf{Wy},
\end{equation}
with $\mathbf{W}$ being
\begin{equation}
    \mathbf{W} = \eta\big(\mathbf{W}_{\text{RF}}^{H}\big)^\dagger = \frac{S\bar{S}}{\text{tr}\Big(\big(\mathbf{W}_{\text{RF}}^{H}\big)^\dagger\mathbf{W}_{\text{RF}}^{H}\Big)}\big(\mathbf{W}_{\text{RF}}^{H}\big)^\dagger.
    \label{eq:pinv}
\end{equation}

Equation~(\ref{eq:pinv}) constructs the linear estimator using only the known combiner $\mathbf{W}_{\text{RF}}$ via the pseudoinverse $\big(\mathbf{W}_{\text{RF}}^{H}\big)^\dagger$ and a scalar $\eta$. Importantly, $\mathbf{W}_{\text{RF}}$ is predefined and does not rely on perfect CSI. Thus, no oracle information about the true channel is used.

The scalar $\eta$ ensures de-correlation by satisfying
\begin{equation}
    \text{tr}\big(\mathbf{I} - \mathbf{W}\mathbf{W}_{\text{RF}}^{H}\big) = 0,
\end{equation}
as discussed in~\cite[Eq. (14)]{2017Ma}. This linear estimate $\mathbf{\hat{h}}_0$ serves as the input to the BRT block at the first iteration, as shown on the left side of \figname~\ref{fig:rbt-model-unrolled}.

During each iteration, the updated channel estimate $\mathbf{\hat{h}}_t$ and internal memory states $\{\mathbf{c}_t^{(l)}\}$ are passed forward to the next step. After $N_t$ iterations, the final output of the model is taken to be
\begin{equation}
    \mathbf{\hat{h}} = \mathbf{\hat{h}}_{N_t},
\end{equation}
which provides the refined channel estimation vector.

Thus, the combination of linear initialization and recurrent BRT-based refinement enables the model to iteratively denoise and enhance the channel estimate, leveraging both vertical expressiveness and horizontal memory dynamics.

\section{Training Procedure and Simulation Setup}
\label{sec:training-eval-setup}

\subsection{Training Setup}
\label{sec:training-setup}

Most supervised machine learning models are trained on datasets of limited size. For instance, computer vision models are trained on sets of images, NLP models are trained on text corpora, and regression models are trained on recorded time series. However, the limited size of these datasets often leads to overfitting, which can be attributed to factors such as noise, insufficient data, and excessive model complexity~\cite{2019Roelofs}. In many real-world scenarios, obtaining labeled data is expensive and time-consuming. To mitigate these challenges, researchers utilize various techniques, including regularization, early stopping, dropouts, data augmentation, and other specialized training methods~\cite{2019Ying, 2022Santos, 2025Zhang}.

In some applications, dataset size is not a limiting factor. For instance, in certain deep learning tasks, models can be trained on synthetic data generated on-the-fly by a mathematical model that describes the system’s properties\footnote{Several studies propose to train digital twins of the wireless channels, which can also be applied for generating the training data, see~\cite{2023Hoydis} for more details.}. This method is commonly used in channel coding~\cite{2025Artemasov, 2022Choukroun}, where generating and storing the datasets for practically relevant code lengths is computationally infeasible due to the curse of dimensionality. A similar approach can also be used in signal processing tasks for wireless communications, as in this paper. The key idea is to generate training data on-the-fly using a stochastic model with predefined properties. When such an approach is employed, the overfitting problem is mitigated since a new set of samples is generated for each training epoch.

In this paper, we utilize the BRT model, which has strong representational power but may be prone to overfitting when trained on limited datasets. To address this issue, we adopt an online dataset generation approach, ensuring that new channel responses $\mathbf{h}$ and measurements $\mathbf{y}$ are generated for every epoch of the training procedure. This means that for each training and validation cycle, new data is produced, effectively preventing overfitting, broadening the distributional coverage, reducing sampling bias, and better matching deployment conditions.

The hybrid-field channel model is defined in (\ref{eq:channel}), with its specific parameters summarized in Table~\ref{tab:channel-parameters}. The training procedure proceeds as follows: for each epoch, $500\,000$ training samples and $50\,000$ validation samples are generated according to equation~(\ref{eq:real-model}) and passed to the BRT model using a dataloader with a batch size of 64. The training is performed over an SNR range of $0$ to $20$~dB. The NMSE, as defined in~(\ref{eq:nmse}), serves as the target metric, i.e., loss function. This metric is minimized using the AdamW optimizer~\cite{2018Loshchilov} with an initial learning rate of $5\cdot10^{-5}$ and a weight decay factor of $0.01$. During training, the learning rate is adjusted using a ``reduce on plateau'' scheduler. The training is performed for $150$ epochs. These hyperparameters are selected because they provide the fastest convergence while maintaining training stability.

\begin{table}[t]
\caption{Channel Simulation Parameters\label{tab:channel-parameters}}
\centering{}
\begin{tabular}{|l|l|}
\hline 
\textbf{Parameter} & \textbf{Value}\tabularnewline
\hline 
Number of RF chains / SAs & $S=4$\tabularnewline
Number of AEs per SA & $\bar{S}=256$\tabularnewline
Carrier frequency & $f_{c}=300$ GHz\tabularnewline
SA spacing & $d_{\text{sub}}=5.6\times10^{-2}$ m\tabularnewline
AE spacing & $d_{a}=5.0\times10^{-4}$ m\tabularnewline
Pilot length & $Q=128$\tabularnewline
Elevation AoA & $\theta_{l}\sim\mathcal{U}(-\pi/2,\pi/2)$\tabularnewline
Azimuth AoA & $\phi_{l}\sim\mathcal{U}(-\pi,\pi)$\tabularnewline
Angle of incidence & $\varphi_{\text{in},l}\sim\mathcal{U}(0,\pi/2)$\tabularnewline
Number of paths & $L=5$\tabularnewline
Rayleigh distance & $Z=20$ m\tabularnewline
LoS path length & $\text{\ensuremath{r_{1}=30}}$ m\tabularnewline
Scatterer distance ($l>1$) & $r_{l}\sim\mathcal{U}(10,25)$ m\tabularnewline
Time delay of LoS path & $\tau_{1}=100$ ns\tabularnewline
Time delay of NLoS paths ($l>1$) & $\tau_{l}\sim\mathcal{U}(100,110)$ ns\tabularnewline
Absorption coefficient & $k_{\text{abs}}=0.0033$ m$^{-1}$\tabularnewline
Refractive index & $n_{t}=2.24-j0.025$\tabularnewline
Roughness factor & $\sigma_{\text{rough}}=8.8\times10^{-5}$ m\tabularnewline
\hline 
\end{tabular}
\end{table}

The described training scheme places high demands on both GPU and CPU performance. The GPU performance determines the speed of model training, while the CPU performance is critical for online dataset generation, which must occur in parallel with the training process. In our setup, dataset generation is handled by two AMD EPYC 7543 processors (128 threads in total), and the model training is performed on an NVIDIA A100 GPU. With this configuration, training the model, with the parameters specified in Table~\ref{tab:model-parameters}, from scratch takes approximately 36 hours. Nevertheless, it is worth noting that, although training the model requires significant computational resources and time, the model remains efficient during inference. We report inference-time performance in Appendix~\ref{app:inference-time} and model sensitivity to hyperparameter values in Appendix~\ref{app:hyper-sensitivity}.

\begin{table}[t]
\caption{Block Recurrent Transformer Model Parameters\label{tab:model-parameters}}
\centering{}%
\begin{tabular}{|l|l|}
\hline 
\textbf{Parameter} & \textbf{Value}\tabularnewline
\hline 
Depth & $N_l=3$\tabularnewline
Hidden dimension & $N_h=2S\bar{S}=2048$\tabularnewline
Number of attention heads & $N_{\text{heads}}=1$\tabularnewline
Dimension of attention heads & $N_{h_{\text{heads}}}=1024$\tabularnewline
Number of recurrent iterations & $N_t=5$\tabularnewline
\hline 
\end{tabular}
\end{table}
\subsection{Evaluation Performance}

The performance of the proposed model is assessed on the baseline hybrid-field channel used in~\cite{2022LetaiefConf, 2022Letaief}. Its parameters are outlined in Table~\ref{tab:channel-parameters}. We benchmark against the following methods:
\begin{itemize}
\item \textbf{LS}: Least squares estimation.
\item \textbf{OMP}: Orthogonal matching pursuit, described in \cite{2021Dovelos}. 
\item \textbf{OAMP}: Orthogonal approximate message passing with pseudoinverse linear estimation, presented in \cite{2017Ma}.
\item \textbf{FISTA}: Fast iterative soft thresholding algorithm, introduced in \cite{2009Beck}.
\item \textbf{EM-GEC}: Bayesian estimation employing expectation-maximization for signal recovery with a Gaussian mixture prior, detailed in \cite{2019Wang}.
\item \textbf{ISTA-Net+}: A deep unfolding technique based on the iterative soft thresholding algorithm, proposed in \cite{2018Zhang}.
\item \textbf{FPN-OAMP}: An advanced method combining classic OAMP with a neural network-based non-linear estimator, outlined in \cite{2022Letaief}. Current state-of-the-art DNN-based solution.
\item\textbf{Conference version}: BRT trained on a fixed \emph{offline} dataset~\cite{artemasov2024sibircon}.
\end{itemize}

\begin{figure}
    \centering
    \includegraphics[width=\columnwidth]{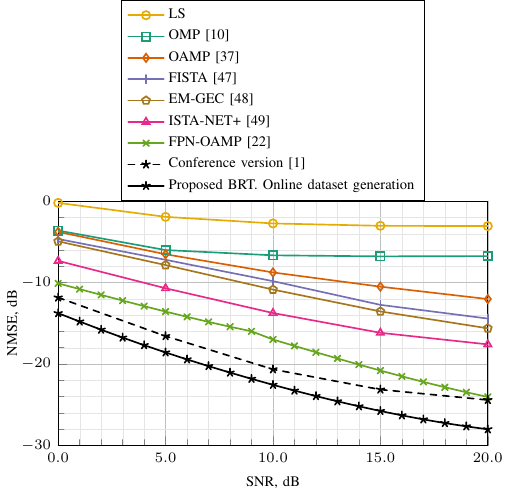}
    \caption{NMSE performance of the hybrid-field channel estimation methods.}
    \label{fig:nmse-results}
\end{figure}

The outcomes in Fig.~\ref{fig:nmse-results} show that the BRT model trained on the dataset that was generated once (offline dataset generation) already exceeds the performance of existing classical and learning-based baselines across the entire SNR range. Moving from the fixed offline dataset to the online dataset generation strategy yields a consistent additional NMSE reduction at all SNRs, making the \emph{Proposed BRT} the top performer overall.\footnote{Our experiments indicate that applying online dataset generation to other DNN-based estimators in the comparison does not provide a noticeable gain. Therefore, we report online-generation results only for the proposed model.}

A small step in the FPN-OAMP curve around $10$~dB in Fig.~\ref{fig:nmse-results} arises because two separate FPN-OAMP models are used, trained for $[0,10]$~dB and $[10,20]$~dB SNR ranges, respectively, as specified in the original repository~\cite{2024YuGithub}. In contrast, BRT model is trained across the full SNR interval, avoiding any SNR-dependent model switching.

\section{Improving the Model's Generalization Ability}
\label{sec:robustness}

\begin{figure}[t]
    \centering
    \includegraphics[width=\columnwidth]{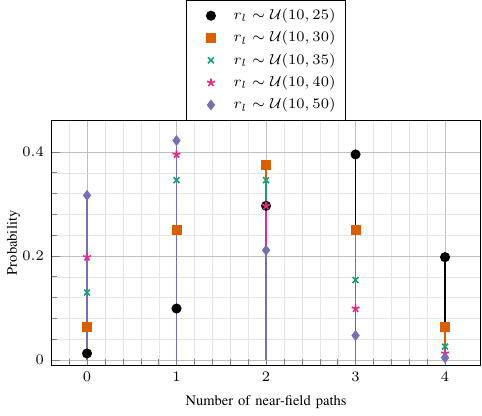}
    \caption{Probability mass function of the number of near-field paths in the channel response for various scatterers distributions $r_l$.}
    \label{fig:nfc-pmf}
\end{figure}

As was mentioned in the introduction, the propagation characteristics of THz waves differ significantly from those in lower frequency bands, such as microwave or even millimeter-wave communications. One critical aspect of THz channel modeling is the number of propagation paths, which is influenced by several factors unique to this frequency range.

The average number of paths in THz channel models is generally limited, in comparison to sub-THz bands, due to the following factors:

\begin{itemize}
    \item \textbf{High Path Loss:} THz signals experience significantly higher free-space path loss compared to lower frequencies, which restricts the number of viable propagation paths. This is primarily due to the inverse relationship between path loss and the square of the wavelength, as described by the Friis transmission equation~\cite{2001Rappaport}.
    \item \textbf{Molecular Absorption:} THz waves are strongly absorbed by atmospheric gases, particularly water vapor. This absorption leads to additional attenuation, further reducing the number of effective propagation paths~\cite{2011Akyildiz}.
    \item \textbf{Scattering and Reflection:} While THz waves can be reflected or scattered by surfaces, the reflectivity and roughness of materials at these frequencies play a crucial role. Smooth surfaces, such as metals, can produce strong reflections, but rough surfaces tend to scatter the signal, reducing the number of dominant paths.
    \item \textbf{Directionality:} To mitigate the high path loss, THz communication systems often employ highly directional antennas. This directionality limits the number of multipath components, as the signal is concentrated in a narrow beam.
\end{itemize}

In practical scenarios, the number of propagation paths in THz channels is typically sparse: in indoor environments, the number of paths is usually limited to a few strong reflections (e.g., 2–5 paths), depending on the room geometry and material properties~\cite{2013Kurner}, while in outdoor scenarios, the channel is often dominated by the line-of-sight (LOS) path, with non-line-of-sight (NLOS) paths being negligible unless strong reflectors (e.g., buildings or walls) are present~\cite{2024Wang}.

In this paper, we investigate the model's ability to generalize across various channel conditions. To maintain conciseness, the performance of the proposed hybrid-field channel estimation model (BRT) is compared only to the previous state-of-the-art DNN-based method, FPN-OAMP \cite{2022Letaief}.

\subsection{Robustness of the DNN-Based Channel Estimation Models to the Ratio of Far- and Near-Field Paths}

\begin{figure}
    \centering
    \includegraphics[width=\columnwidth]{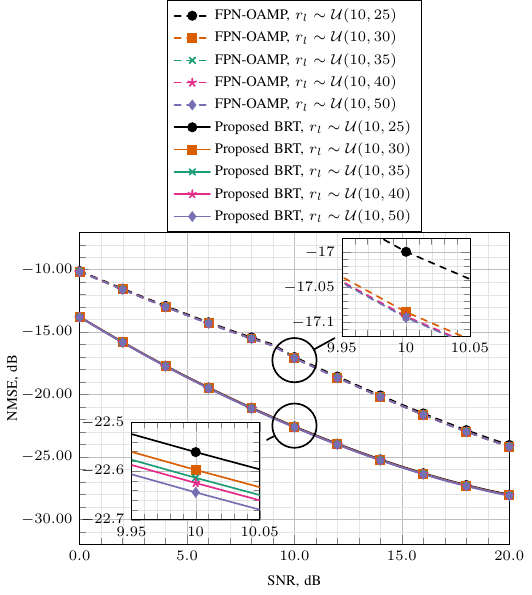}
    \caption{NMSE performance across channels with various scatterer distributions.}
    \label{fig:delta-nmse-distance}
\end{figure}

In the baseline scenario (see Table~\ref{tab:channel-parameters}), the LoS path length $r_1$ is set to 30 meters, which corresponds to the far-field propagation. The distances of the NLoS paths (scatterers) are uniformly distributed between 10 and 25 meters. To validate the robustness of the model, we evaluate its performance on channels with different scatterer distance ranges. In other words, we evaluate the channel estimation performance of the model pretrained on a channel with $r_{l}\sim\mathcal{U}(10,25)$ (as described in Section~\ref{sec:training-eval-setup}) on channels with other $r_l$ distributions.

Using the $r_l$ distribution and noting that the number of paths is fixed $L=5$, we can calculate the probability of a channel realization containing a specific number of far-field and near-field paths. \figname~\ref{fig:nfc-pmf} depicts the probability mass function (PMF) of the number of near-field paths in the channel response as a function of the scatterer distance limits.

The model's performance is presented in \figname~\ref{fig:delta-nmse-distance}, where it is compared with FPN-OAMP from \cite{2022Letaief}. From the figure, it can be observed that both FPN-OAMP and BRT models trained on the baseline channel demonstrate strong generalization capabilities for channels with an increased maximum distance of scatterer distribution. The NMSE performance on the analyzed channels is slightly better than that on the baseline channel. This can be attributed to the fact that increasing the maximum distance of scatterer distribution enhances the prevalence of planar waves in the channel response, and the power of NLoS paths decreases due to propagation losses.

\subsection{Robustness of the DNN-Based Channel Estimation Models to the Number of the Propagation Paths}

\begin{figure}
    \centering
    \includegraphics[width=\columnwidth]{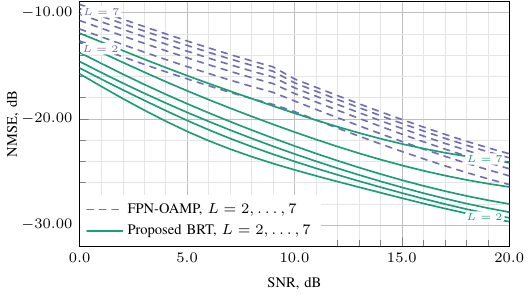}
    \caption{NMSE performance across channels with varying number of propagation paths $L$. Models trained on the baseline channel ($L=5$).}
    \label{fig:nmse-5-5}
\end{figure}

\begin{figure}
    \centering
    \includegraphics[width=\columnwidth]{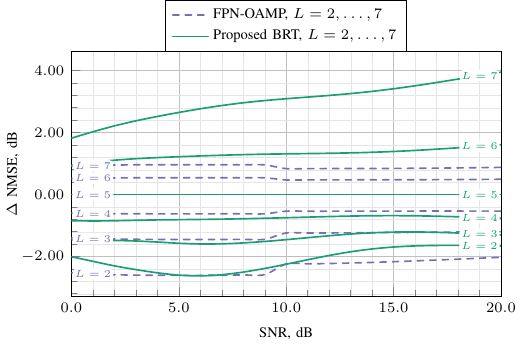}
    \caption{Difference in NMSE performance across channels with varying number of propagation paths $L$. Models trained on the baseline channel ($L=5$).}
    \label{fig:delta-nmse-5-5}
\end{figure}

We now experimentally analyze the generalization ability of the proposed model (BRT) and the baseline model FPN-OAMP to different numbers of propagation paths $L$ in the channel. We begin with models trained on a baseline channel, when $L=5$ (see Table~\ref{tab:channel-parameters}), and evaluate their performance on channels with $L=2,\dots  ,7$. \figname~\ref{fig:nmse-5-5} presents the NMSE performance of both models for $L=2, \dots , 7$. Additionally, \figname~\ref{fig:delta-nmse-5-5} illustrates the NMSE performance difference between models validated on channels with varying $L$ values and those trained and validated on the baseline channel. The results indicate that both models perform better on channels with fewer paths ($L=2,\dots,4$) and experience performance degradation as $L$ increases ($L=6,7$). However, the proposed BRT model consistently outperforms the baseline.

\begin{figure}
    \centering
    \includegraphics[width=\columnwidth]{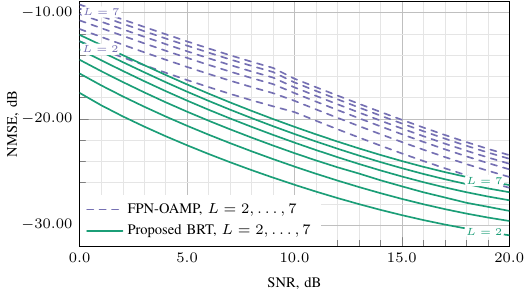}
    \caption{NMSE performance across channels with varying number of propagation paths $L$. Models trained on the dataset with $L\sim\mathcal{U}\{2,7\}$.}
    \label{fig:nmse-2-7}
\end{figure}

\begin{figure}
    \vspace{4px}
    \centering
    \includegraphics[width=\columnwidth]{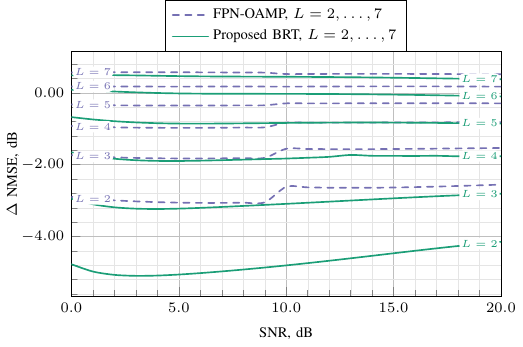}
    \caption{Difference in NMSE performance across channels with varying number of propagation paths $L$. Models trained on the channel with number of paths $L\sim\mathcal{U}\{2,7\}$.}
    \label{fig:delta-nmse-2-7}
\end{figure}

To enhance the robustness of the proposed model under varying channel conditions, we propose a different training strategy. Specifically, we train the proposed model on channels where the number of paths $L$ varies within each epoch. To achieve this, during training, the value of $L$ is sampled from a discrete uniform distribution independently for each training sample, ${L \sim \mathcal{U}\{2,7\}}$, instead of a fixed $L=5$. The NMSE results for models trained under this setting are shown in \figname~\ref{fig:nmse-2-7}, while \figname~\ref{fig:delta-nmse-2-7} depicts the NMSE differences across various channel conditions. The results demonstrate that this training approach does not introduce significant performance loss compared to the baseline, while improving the model’s robustness and generalization to different channel conditions.

\section{Wideband Channel Estimation}
\label{sec:wideband}

To achieve the throughput capabilities of the IMT-2030 (6G) defined in~\cite{2024ITU-IMT2030}, the width of the utilized frequency bands in the THz ranges should be substantial. The need for throughput capacity of up to 1 Tbit/s requires the aggregation of additional spectrum, specifically the 2–3 GHz range for widespread deployment, and tens of GHz for localized solutions~\cite{2024ETSIThzBands}. THz and sub-THz frequencies are essential due to their abundant channel availability. However, their use requires compensation for propagation loss, as mentioned in the introduction section. In addition to the significant signal absorption at THz frequencies, wideband communication systems also experience the beam squint effect~\cite[section~5.4]{2025ETSIThzModeling},~\cite{2023Wanming}, which introduces further challenges.

Let us consider a wideband THz OFDM system with $K$ subcarriers and bandwidth $B$. Recalling equation (\ref{eq:real-model}), we can reformulate it to model the wideband channel as follows
\begin{equation}
    \mathbf{y}_k = \mathbf{W}_{\text{RF}}^H \mathbf{h}_k + \mathbf{n}_k\quad\forall k\in[1,\dots,K].
    \label{eq:real-model-wideband}
\end{equation}

In \eqref{eq:real-model-wideband}, $\mathbf{y}_k$, $\mathbf{h}_k$ and $\mathbf{n}_k$ are the received signal, channel response vector, and the AWGN, respectively, at the subcarrier frequency
\begin{equation}
    {f_k=f_c + \left(k - 1 - \frac{K-1}{2}\right)\frac{B}{K}}.
\end{equation}

We assume that the pilot signals transmission occurs simultaneously over all $K$ subcarriers, so the measurement matrix $\mathbf{W}_{\text{RF}}^H$ remains the same for all subcarriers. Also, as was mentioned earlier, due to the high carrier frequency and large bandwidths of THz communication systems, different subcarriers may experience different propagation losses. To account for this, we calculate the molecular absorption loss coefficients $k_{abs}$, used in equation (\ref{eq:los-path-loss}), independently for each subcarrier in accordance with the HITRAN model~\cite{2011Akyildiz, 2021Tarboush}.

In our simulations, we follow the wideband setup described in~\cite{2022Letaief}. Specifically, we extend the scenario defined in Table~\ref{tab:channel-parameters} with a bandwidth of $B=15$ GHz and $K=32$ subcarriers. To adapt to this wideband channel setting, both the proposed and baseline channel estimation models were fine-tuned on wideband data. For the FPN-OAMP model, the subcarrier dimension was treated as a separate sample dimension during fine-tuning, effectively increasing the batch size by a factor of $K$, as proposed in~\cite{2022Letaief}.

One of the key contributions of this work lies in how the channel estimation model is applied to the wideband scenario. The proposed architecture explicitly captures inter-subcarrier dependencies. Unlike methods such as FPN-OAMP, which treat each subcarrier independently by processing per-subcarrier channel estimation vectors in isolation, our model processes the channel estimation as a wideband estimation matrix, $\mathbf{\hat{H}}_t \in \mathbb{R}^{K \times 2S\bar{S}}$. We treat the subcarrier dimension as a sequence of input tokens, enabling multi-head self-attention to capture spectral dependencies within each subcarrier block. For each block, the attention mechanism first projects the input embeddings into query and key $\mathbf{Q},\mathbf{K}\in \mathbb{R}^{L\times d_k}$ matrices via learned linear transformations. Then it computes a square similarity matrix $\mathbf{QK}^T$ of size equal to the number of subcarriers in the block, where each element quantifies the similarity between a pair of subcarriers based on the dot product of their query and key representations. After softmax normalization, this matrix guides how information is aggregated across correlated frequencies via the weighted sum of values. Blocks are further linked by a recurrent state that propagates contextual information across blocks, preserving long-range frequency correlations. This hierarchical combination of local attention and inter-block recurrence allows the model to exploit both fine-scale and global subcarrier correlations, which is expected to significantly enhance wideband signal recovery.

\begin{figure}
   \centering
   \includegraphics[width=\columnwidth]{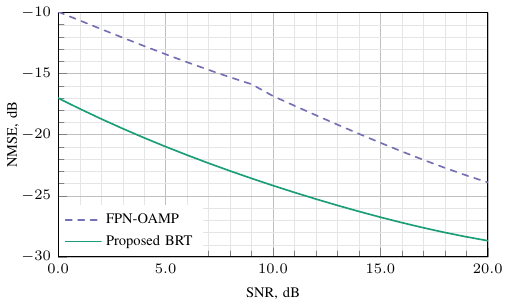}
   \caption{NMSE performance of the wideband hybrid-field channel estimation methods.}
   \label{fig:nmse-results-wideband}
\end{figure}

\begin{figure}
   \centering
   \includegraphics[width=\columnwidth]{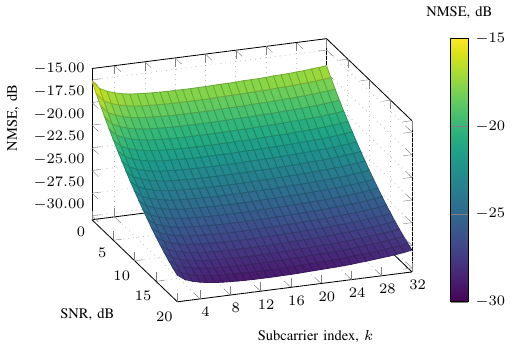}
   \caption{BRT wideband hybrid-field channel estimation NMSE performance.}
   \label{fig:nmse-results-wideband-surf}
\end{figure}

\figname~\ref{fig:nmse-results-wideband} illustrates the NMSE performance of the proposed BRT and FPN-OAMP channel estimation methods. It is evident that the proposed method significantly outperforms FPN-OAMP across the considered SNR range. Additionally, the performance of the proposed method in the wideband scenario is better than in the baseline narrowband scenario, especially at low SNR values (see~\figname~\ref{fig:nmse-results}). Also, \figname~\ref{fig:nmse-results-wideband-surf} presents the surf plot illustrating the performance of the proposed model across $K$ subcarriers at various SNR levels. The results indicate that the performance does not change much across different subcarriers, demonstrating the strong generalization ability of the model over the entire considered band. We provide additional analysis of the model's sensitivity to bandwidth/beam-squint effects in Appendix~\ref{app:beam-squint}.

\section{Conclusions}
\label{sec:conclusions}

In this paper, we proposed a novel channel estimation method based on the block recurrent transformer architecture. We also proposed strategies for training the model in order to improve its generalization ability to different channels and bandwidth scenarios. Experimental results demonstrate that the proposed model consistently and significantly outperforms state-of-the-art solutions across all evaluated scenarios. The method was tested on channels with varying parameters, including different scatterer distribution distances, numbers of propagation paths, and in wideband conditions. Notably, a model with a single set of trainable weights achieved superior performance across a $20$~dB SNR range, highlighting its robustness and adaptability. Although presented in the THz hybrid-field wideband setting, the proposed BRT architecture is general and can be retrained for other bands and for purely far-field models without changing the core design. We highlight the analysis and mitigation of the model's computational complexity as an important avenue for future research.

\appendices
\section{Model Inference Time Analysis}
\label{app:inference-time}

\begin{table*}[t]
\caption{Inference Time of BRT and FPN-OAMP Models\label{tab:inference-time}}
\centering
\begin{tabular}{|c|c|c|c|c|c|}
\hline 
\makecell{\textbf{Number of} \\ \textbf{subcarries}} & 
\makecell{\textbf{Batch} \\ \textbf{size}} & 
\makecell{\textbf{BRT forward pass} \\ \textbf{time per batch, ms}} & 
\makecell{\textbf{FPN-OAMP forward pass} \\ \textbf{time per batch, ms}} & 
\makecell{\textbf{BRT forward pass} \\ \textbf{time per sample, ms}} & 
\makecell{\textbf{FPN-OAMP forward pass} \\ \textbf{time per sample, ms}} \\
\hline 
1 & 1 & 20 & 12.6 & 20 & 12.6\tabularnewline
& 2 & 20.6 & 13.2 & 10.3 & 6.6\tabularnewline
& 4 & 20.7 & 13.9 & 5.17 & 3.48\tabularnewline
& 8 & 20.8 & 14.5 & 2.6 & 1.8\tabularnewline
& 16 & 20.9 & 14.6 & 1.3 & 0.91\tabularnewline
& 32 & 20.7 & 14.3 & 0.65 & 0.45\tabularnewline
& 64 & 20.6 & 14.8 & 0.322 & 0.23\tabularnewline
& 128 & 20.5 & 20 & 0.16 & 0.157\tabularnewline
& 256 & 33.8 & 33.8 & 0.132 & 0.132\tabularnewline
& 512 & 59.6 & 94.1 & 0.116 & 0.184\tabularnewline
\hline
32 & 1 & 21 & 13.9 & 21 & 13.9\tabularnewline
& 2 & 20.9 & 13.9 & 10.4 & 6.93\tabularnewline
& 4 & 20.6 & 19.9 & 5.15 & 4.97\tabularnewline
& 8 & 33.9 & 33.6 & 4.2 & 4.2\tabularnewline
& 16 & 59.2 & 67.6 & 3.7 & 4.23\tabularnewline
& 32 & 110 & 130.7 & 3.4 & 4.09\tabularnewline
& 64 & 208 & 257 & 3.2 & 4\tabularnewline
& 128 & 414 & 511 & 3.2 & 3.99\tabularnewline
& 256 & 810 & 1022 & 3.16 & 3.99\tabularnewline
& 512 & 1604 & 2042 & 3.13 & 3.99\tabularnewline
\hline 
\end{tabular}
\end{table*}

This appendix presents an analysis of the inference time for the two evaluated models: the current state-of-the-art FPN-OAMP~\cite{2022Letaief} (used as the baseline) and the proposed BRT model.

For this comparison, we employed the publicly available implementation of FPN-OAMP from~\cite{2024YuGithub}, using the default parameters specified in~\cite{2022Letaief, 2024YuGithub}. The parameters of the proposed BRT model are detailed in Table~\ref{tab:model-parameters}. Inference for both models was conducted using single-precision floating-point arithmetic on the hardware configuration described in Section~\ref{sec:training-setup}.

Table~\ref{tab:inference-time} presents the inference time for both the proposed BRT model and the baseline FPN-OAMP model across varying batch sizes and subcarrier configurations. Overall, the BRT model outperforms FPN-OAMP in terms of per-sample inference time, at larger batch sizes. For instance, in a narrowband scenario with a batch size of 512, BRT achieves a per-sample inference time of 0.116 ms, compared to 0.184 ms for FPN-OAMP. The performance gap becomes even more pronounced when the number of subcarriers increases: with 32 subcarriers and a batch size of 512, BRT requires 3.13 ms per sample, whereas FPN-OAMP takes 3.99 ms.

Taking into account the demonstrated channel estimation performance of the BRT model, it becomes evident that BRT offers a favorable trade-off between accuracy and computational cost. Its significantly lower per-sample inference time, particularly at larger batch sizes and higher subcarrier configurations, positions it as a strong candidate for large-scale deployment scenarios. Furthermore, the reported results are based on single-precision floating-point inference. Additional reductions in inference time may be achievable through quantization or the use of lower bitwidth representations, which we leave for future investigation.

\section{Hyperparameter Sensitivity}
\label{app:hyper-sensitivity}

In this section, we investigate the sensitivity of the model to its hyperparameter values. In particular, we examine how the number of recurrent iterations $N_t$, the number of attention heads $N_\text{heads}$, and the model depth $N_l$ affect the final NMSE channel estimation performance.

\begin{figure}[t]
    \centering
    \includegraphics[width=\columnwidth]{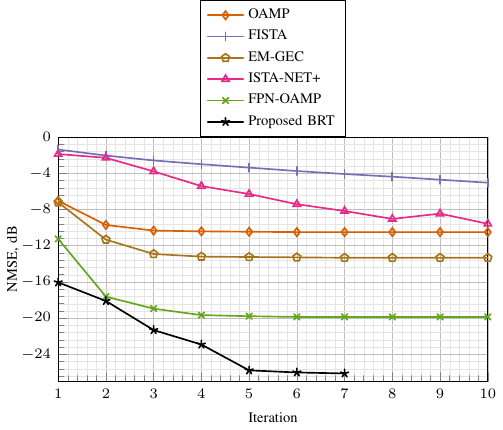}
    \caption{Effect of the number of recurrent iterations $N_t$ on NMSE performance at $15$~dB SNR, including baselines from~\cite{2022Letaief}. Identical settings (Table~\ref{tab:channel-parameters}) ensure comparability.
    }
    \label{fig:time-steps}
\end{figure}

\figname~\ref{fig:time-steps} reports the NMSE achieved by the proposed BRT model in comparison with baselines from~\cite{2022Letaief} for ${N_t=1,\ldots,10}$ recurrent iterations at the SNR value of $15$~dB. All other hyperparameters remain unchanged from Table~\ref{tab:model-parameters}. The experiments follow the reference simulation setup specified in Table~\ref{tab:channel-parameters}.

The results indicate that the model performance improves as $N_t$ increases. In the considered setting, most of the performance gain of the proposed model is realized within the first five iterations, after which the performance improvement saturates. However, the obtained results demonstrate a faster convergence rate and better channel estimation accuracy compared to competing methods.

We emphasize that the optimal number of refinement iterations depends on the characteristics of the underlying system model: for example, simple LoS channels typically require fewer iterations than rich multipath environments. Overall, the proposed architecture is universal and can be applied to MIMO channel estimation tasks across a wide range of scenarios, with specific hyperparameter choices tailored to the application requirements.

Table~\ref{tab:attention-heads} provides a performance comparison of the proposed model trained with different numbers of attention heads $N_\text{heads}$. In this set of experiments, $N_\text{heads}$ is varied while all other hyperparameters remain fixed to the values listed in Table~\ref{tab:model-parameters}. As shown, the estimation accuracy improves as the number of attention heads increases. However, selecting an appropriate number of heads represents a trade-off between accuracy and computational complexity, and the optimal choice may depend on the channel environment and model capacity constraints.

\begin{table}[t]
\centering
\caption{Performance comparison for different numbers of attention heads $N_\text{heads}$ at $15$~dB SNR. Boldface indicates results for the default hyperparameter setting (Table~\ref{tab:model-parameters}).}
\label{tab:attention-heads}
\begin{tabular}{|c|c|}
\hline
\makecell{$N_\text{heads}$} & \makecell{NMSE [dB]} \\
\hline
$\mathbf{1}$ & $\mathbf{-25.77}$ \\
$2$ & $-26.24$ \\
$4$ & $-27.13$ \\
\hline
\end{tabular}
\end{table}

Table~\ref{tab:model-depth} provides a performance comparison of the proposed model trained with different network depths ${N_l=1,\ldots,4}$. In this experiment, the model depth is varied while all other hyperparameters are fixed. As the results show, increasing the model depth generally improves the channel estimation accuracy, demonstrating that deeper architectures can better capture the underlying channel structure. However, the performance gains diminish beyond a certain depth, indicating that additional layers provide only marginal benefit relative to the increased computational cost and training complexity. Consequently, selecting an appropriate model depth involves balancing estimation accuracy and model efficiency.

\begin{table}[t]
\centering
\caption{Performance comparison for different model depths $N_l$ at $15$~dB SNR. Boldface indicates results for the default hyperparameter setting (Table~\ref{tab:model-parameters}).}
\label{tab:model-depth}
\begin{tabular}{|c|c|}
\hline
\makecell{$N_l$} & \makecell{NMSE [dB]} \\
\hline
$1$ & $-24.34$ \\
$2$ & $-25.47$ \\
$\mathbf{3}$ & $\mathbf{-25.77}$ \\
$4$ & $-25.82$ \\
\hline
\end{tabular}
\end{table}

\section{Bandwidth and Beam-Squint Sensitivity}
\label{app:beam-squint}

At large bandwidths,  beam squint may misalign the analog beams with the frequency-dependent array response, effectively increasing frequency selectivity and inter-subcarrier coupling.

To quantify this effect, we conduct a bandwidth-sensitivity study (all results reported at $15$~dB SNR). We evaluate NMSE across bandwidths $B\in \{5, 10, 15, 20, 25\}$~GHz while keeping the center frequency, pilot/combiner design, and total pilot budget fixed. In the first experiment, the model is trained only in the reference setup ($B=15$~GHz with $K=32$ subcarriers), which serves as the wideband baseline used in the paper. The results are present in Table~\ref{tab:beam-squint} under ``w/o fine-tuning'' column. The second experiment (Table~\ref{tab:beam-squint}, ``with fine-tuning'') starts from the model trained on $B=15$~GHz and fine-tunes it for each target bandwidth.

From Table~\ref{tab:beam-squint}, the fine-tuned model yields small gains at lower bandwidths and negligible changes at higher bandwidths. The non-fine-tuned model performs best at its training bandwidth ($B=15$~GHz) and shows NMSE degradation of up to $5$~dB when evaluated at other bandwidths. We attribute the generally strong performance to the model's subcarrier-tokenized attention, which leverages inter-subcarrier correlation and mitigates squint-induced distortions. In practice, systems typically operate at a fixed bandwidth, for which the model's trainable parameters can be optimized. However, if we consider a communication system, in which the bandwidth can change during the operation, a small bank of bandwidth-specific fine-tuned weights can be used to retain near-optimal performance.

\begin{table}[h]
  \centering
  \caption{Bandwidth sensitivity of the BRT-based channel estimator at $15$~dB SNR. 
  Number of subcarriers fixed at $K=32$, center frequency, pilot/combiner design, and pilot budget held constant. Boldface indicates results for the default bandwidth setting.}
  \label{tab:beam-squint}
  \begin{tabular}{|l|c|c|}
    \hline
    Bandwidth & \makecell{w/o fine-tuning,\\ NMSE [dB]} & \makecell{with fine-tuning,\\ NMSE [dB]}\\
    \hline
    $5$~GHz  & $-21.54$ & $-27.25$ \\
    $10$~GHz & $-24.95$ & $-26.99$ \\
    $\mathbf{15}$~\textbf{GHz} & $\mathbf{-27.08}$ & $\mathbf{-27.08}$ \\
    $20$~GHz & $-25.17$ & $-26.98$ \\
    $25$~GHz & $-22.98$ & $-26.80$ \\
    \hline
  \end{tabular}
\end{table}

\bibliographystyle{IEEEtran}
\bibliography{refs}

\end{document}